\documentclass[useAMS,usenatbib]{mnras}
\usepackage{graphicx}
\usepackage{txfonts}
\def\aap{A\&A}
\def\apj{ApJ}
\def\apjl{ApJL}
\def\mnras{MNRAS}

\title[Oscillation-induced instabilities in jets]{Propagation, cocoon formation, and resultant destabilization of relativistic jets}
\author[J. Matsumoto and Y. Masada]{Jin Matsumoto$^{1}$
\thanks{E-mail: jin.matsumoto@fukuoka-u.ac.jp, jin@kusastro.kyoto-u.ac.jp} and Youhei Masada$^{2}$ \\
$^{1}$Research Institute of Stellar Explosive Phenomena, Fukuoka University, Fukuoka 814-0180, Japan\\
$^{2}$Department of Science Education, Aichi University of Education, Kariya 448-8542, Japan}
\begin{document}
\date{Accepted xxx, Received yyy, in original form zzz}
%\date{Accepted 2010 November 9. Received 2010 November 2; in original form 2010 October 2}
\pagerange{\pageref{firstpage}--\pageref{lastpage}} \pubyear{2019}
\maketitle
\label{firstpage}
\begin{abstract}
A cocoon is a by-product of a propagating jet that results from shock
heating at the jet head. Herein, considering simultaneous cocoon formation, 
we study the stability of relativistic jets propagating through the uniform 
ambient medium. Using a simple analytic argument, we demonstrate that 
independent from the jet launching condition, the effective inertia of the 
jet is larger than that of the cocoon when the fully relativistic jet oscillates
radially owing to the pressure mismatch between jet and cocoon.
In such situations, it is expected that the onset condition for the 
oscillation-induced Rayleigh--Taylor instability is satisfied at the jet interface,
resulting in the destabilization of the relativistic jet during its propagation.
We have quantitatively verified and confirmed our prior expectation by performing 
relativistic hydrodynamic simulations in three dimensions. The possible
occurrences of the Richtmyer--Meshkov instability, oscillation-induced
centrifugal instability and Kelvin--Helmholtz instability are also discussed.
\end{abstract}


\section{Introduction}
One of characteristic features of the astrophysical jet is its coherency
in space and time. In contrast to that, it is implied from the recent
studies that, in the relativistically-propagating jet, there exists
small-scale disturbances, i.e., turbulence, which plays a considerable
role in the particle acceleration and flaring activities of the jetted flow
\citep[e.g.,][]{Asano15b, Asano15a, Asano18}. The large-scale coherency
and small-scale incoherency, i.e., driven turbulence, are both fundamentally
related to the nonlinear stability of the relativistic jet. 

The evolution of the relativistic jet propagating through an ambient 
medium has been studied in high energy astrophysics. The dynamics
and stability of jets from active galactic nuclei (AGNs) have been
investigated through relativistic numerical simulations from 1990s 
\citep{Marti97, Gomez97, Komissarov97}. The propagation of the
relativistic jet drilling a massive star is a key process in order to
reveal the origin of the radiation mechanism for long gamma-ray
bursts \citep[GRBs;][]{Aloy00, Zhang03, Mizuta06, Morsony07,
Lazzati09, Nagakura11, Ito15, Ito19, Gottlieb19}. %Mizuta09
The dynamics of a short GRB jet associated with a compact binary
merger is a front line topic to understand electromagnetic counterparts
to gravitational wave signals from the binary system \citep{Aloy05,
Nagakura14, Murguia-Berthier14, Just16, Gottlieb18a, Gottlieb18b}.
The stability of the propagating jet is crucial in order to maintain
coherent structures of these relativistic jets. 

On the other hand, relativistic jets are subjected to a storm of
hydrodynamic (HD) and/or magnetohydrodynamic (MHD) instabilities
when they propagate through the ambient medium. A velocity shear
layer between the jet and surrounding medium is unstable to
Kelvin--Helmholtz instability \citep[KHI;][]{Turland76, Blandford76}.
The observational structures, such as helices of jets, have been
interpreted as the result of the growth of helical modes of the KHI.
A helical motion of the jet itself leads to the deformation of the jet
structure \citep[e.g.,][]{Aloy99, Perucho19}.

The Poynting flux-dominated jets carrying large-scale helical magnetic
fields can become unstable to current driven kink instability
\citep[CDI;][]{Lundquist51, Spruit97, Begelman98}. The growth of
the CDI is also the possible origin of the helical structure of the jet.
In addition, the growth of the CDI may contribute to the energy
conversion of the jet from the magnetic energy into the thermal
energy via the magnetic reconnection \citep{Bromberg16}.

The rotation of the jet, which is the azimuthal component of the velocity
against the jet axial velocity, is also a possible origin of the distortion
of the jet interface. The rotational shears between the different jet
components and/or surrounding medium can become unstable by
the centrifugal-buoyancy force \citep{Meliani07, Meliani09, Millas17}.

Even without the rotation of the jet, the centrifugal force can impact
on the jet stability. \citet{Gourgouliatos18a} reported that the centrifugal
instability (CFI) grew at the interface of AGN jets undergoing the
reconfinement. This instability was responsible for the transition from
the laminar to the turbulent flow at the reconfinement point and may
be related to the physical origin of a morphological dichotomy of the
AGN jets in the Fanaroff-Riley classification \citep{Fanaroff74}.

\citet[][hereafter MM13]{Matsumoto13} has shown
other possible existence of the destabilization of the relativistic jet interface 
via oscillation-induced Rayleigh--Taylor instability (RTI) and associated 
Richtmyer--Meshkov instability (RMI) during jet propagation. Since they
have non-axisymmetric nature, we had not captured them by conventional
axisymmetric simulations of jet propagation \citep[e.g.,][]{Marti97, Gomez97,
Komissarov97, Komissarov98, Scheck02, Mizuta04, Perucho07b, Meliani08,
Mimica09, Mizuta10, Walg13, Mizuta13, Perucho14} while they may have
been excited in three-dimensional (3D) simulations \citep[e.g.,][]{Aloy99,
Aloy03, Hughes02, Zhang04, Rossi08, Lopez-Camara13, Li18, Gottlieb18a,
Gottlieb18b, Gottlieb19, Perucho19}.

The physics of oscillation-induced RTI and RMI during jet propagation
is summarized as follows (see MM13 for details): The radial 
inertia force, which is a main driver of the jet's non-axisymmetric 
evolution, naturally arises when pressure mismatch 
exists between the jet and surrounding medium. Further, this force induces 
the radially oscillating motion of the jet and simultaneously excites 
the RTI at the interface between the jet and surrounding medium. 
Because the pressure mismatch is not alleviated immediately without 
some damping processes, it repeatedly excites the reconfinement 
shock inside the jet \citep{Sanders83, Daly88, Matsumoto12}. 
Then, RMI is additionally excited when the reconfinement 
shock collides with the jet interface and thus exhibits episodic growth 
with each collision \cite[see, e.g.,][for a review on RMI]{Nishihara10}.

The linear theory of RTI at a discontinuous surface of relativistic 
flows is addressed in \citet{Matsumoto17}. The onset condition is
given analytically by
\begin{eqnarray}
\gamma_1^2 \rho_1 h_1^{\prime} > \gamma_2^2 \rho_2 h_2^{\prime} \;,
\label{eq: onset condition}
\end{eqnarray}
where 
\begin{eqnarray}
h^{\prime}:=1+\frac{\Gamma^2}{\Gamma -1}\frac{P_0}{\rho c^2} \;
\end{eqnarray}
and the subscripts $1$ and $2$ represent the physical variables in the 
upper and lower region against the acceleration in the equilibrium 
state, respectively. Herein, $\gamma$ is the Lorentz factor, $\rho$ is 
the rest-mass density, $\Gamma$ is the ratio of specific heats, $c$ 
is the speed of light and $P_0$ is the interface pressure. 
This condition is applicable for analysing the stability of the interface 
among various jet components \citep{Toma17} and between the 
jet and surrounding medium. In the context of jet propagation, the onset
condition ($\ref{eq: onset condition}$) roughly indicates that the
oscillation-induced RTI grows when the jet is effectively heavier than
the surrounding medium. 

When the jet is effectively lighter than the surroundings, the RTI is
definitely stabilized by the negative buoyancy acting as the restoring
force. It may be noteworthy that, in the situation where the RTI stable,
the growth rates of instabilities accompanied by the radial displacement
of the fluid parcel, such as CFI and KHI, would be more or less reduced
by the negative buoyancy because it also acts against their destabilizing
forces. 

In this paper, we focus on the stability of the jet confined by a cocoon,
as seen in the Fanaroff-Riley class II jet or the GRB jet drilling a
progenitor star. Therefore, in the following, subscripts 1 and 2 in
equation (\ref{eq: onset condition}) correspond to jet and cocoon,
respectively.

\begin{figure}
\begin{center}
\scalebox{1}{{\includegraphics{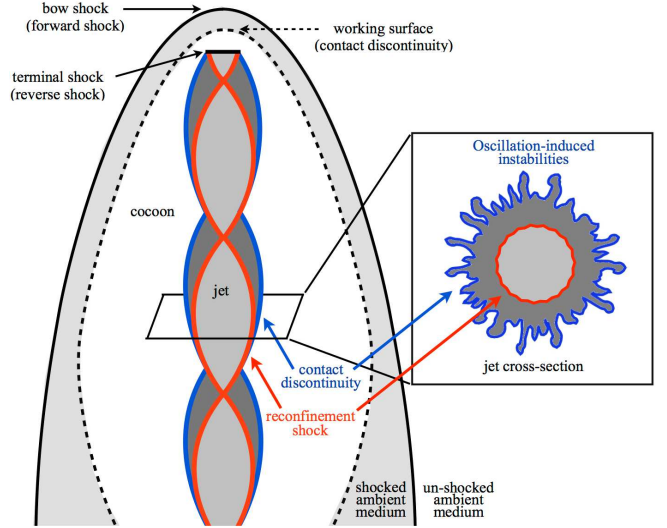}}} 
\caption{Schematic view of jet propagation through the ambient medium.}
\label{fig1}
\end{center}
\end{figure}

The cocoon is generated as a by-product of the jet propagation.
When considering the propagation of a relativistic jet through the 
ambient medium, forward shock (bow shock), reverse shock 
(terminal shock) and contact discontinuity (working surface) are 
formed at the jet head. The matter that enters the jet head 
through forward or reverse shock is heated and separated 
by the working surface. The shocked jet matter forms a hot cocoon 
surrounding the jet itself. (See the schematic of the typical 
jet--cocoon--ambient medium system formed during the jet propagation 
shown in Fig.~\ref{fig1}.) The physical condition of the cocoon
should be dynamically determined by the jet propagation 
\citep{Begelman89, Bromberg11} and thus cannot be understood
without solving the interaction between the jet and ambient medium.
Therefore, in the actual jet--cocoon system, it is unclear whether 
the onset condition for the oscillation-induced RTI is satisfied. 

In this paper, in order to understand the basic physics of the
growth of the oscillation-induced RTI and RMI in the jet, the
nonlinear stability of the relativistic jet propagating through the
uniform ambient medium is addressed by taking account of the 
simultaneous cocoon formation. In Section~2, using a simple
analytic argument, we demonstrate that the condition for the
onset of RTI is possibly satisfied even in the realistic jet-cocoon
system. In Section~3, we test and confirm the analytic prediction
through 3D special relativistic hydrodynamic (SRHD) simulations.
We discuss the possible development of instabilities in jets and
the impact of the decaying pressure/density ambient medium and
the magnetic field on the oscillation-induced RTI in Section~4.
Finally, we summarize our findings in Section 5.
%%%%%%%%%%%%%%%%%%%%%%%%%%%%%%%%%%%%%%%%%%%%%%%%%%%%%
%%%%%%%%%%%%%%%----------S2  --------%%%%%%%%%%%%%%%
%%%%%%%%%%%%%%%%%%%%%%%%%%%%%%%%%%%%%%%%%%%%%%%%%%%%%

%%%%%%%%%%%%%%%%%%%%--------------------Figure 1---------------------------%%%%%%%%%%%%%%%%%
%%%%%%%%%%%%%%%%%%%%%%%%%%%%%%%%%%%%%%%%%%%%%%%%%%%%%%%%%%%

\section{Analytic Prediction on Stability of Relativistic Jet Interface}
Generally, as described in Section~1, the jet is surrounded by
the cocoon formed as a result of the shock heating at the jet head
during its propagation. It is thus expected that oscillation-induced
RTI is excited at the jet interface when the onset condition
(\ref{eq: onset condition}) is satisfied between the jet and cocoon
media. In this section, we demonstrate, with a simple analytic model,
that the onset condition for the RTI is satisfied there as long as
the jet is fully relativistic. 

We suppose a situation where a relativistic flow is continuously
injected into a homogeneous ambient medium. The effects of
the magnetic field and rotation around the jet axis are ignored
here. In such a situation, the shocked ambient medium, cocoon
and jet regions are simultaneously formed inside the lobe of the
forward shock as illustrated in Fig.~\ref{fig1}. The gaseous medium
consisting of each region is assumed to be uniform for the simplicity
in the following.

The recurrence between overexpansion and overcontraction
stages of the jet, which is induced by the pressure mismatch
between the jet and cocoon, results in the radial oscillating motion
of it. Then, time averagely, the pressure of the jet coincides with
that of the cocoon due to the confinement of the jet by the cocoon.
The internal energy and mass of the cocoon garnered through the 
reverse shock during the jet injection time, $t$, thus are estimated,
when all the physical quantities characterizing the jet, such as jet
radius, density, pressure, velocity, and Lorentz factor, are treated as
the ``temporally averaged'' values, as follows: 
\begin{eqnarray}
\frac{P_c}{\Gamma - 1} V_c  = \alpha \pi r_j^2 \gamma_j \rho_j c^2 (\gamma_j h_j - 1)(v_j - v_h)t \;, 
\label{eq: internal energy of cocoon}
\end{eqnarray}
\begin{eqnarray}
M_c = \pi r_j^2 \gamma_j \rho_j (v_j - v_h)t
\label{eq: total mass of cocoon}
\end{eqnarray}
\citep{Marti97, Bromberg11, Perucho17}, where subscripts $j$
and $c$ stand for the jet and cocoon respectively. Here, $V_c$
is the volume of the cocoon, $r_j$ is the jet radius, 
$h := 1 + \Gamma P/(\Gamma-1)\rho c^2$ is the dimensionless specific
enthalpy and $v_h$ is the propagation velocity of the jet head.
The parameter $\alpha$ is the conversion factor from the jet energy
to the internal energy of the cocoon at the jet head. The other symbols
retain the same meanings as those in Section~1.

It was reported from recent numerical studies of the propagation
of AGN jets that roughly $40$\% of the jet energy is converted into
the internal energy of the cocoon medium \citep{Perucho17}. We thus
choose the conversion factor as $\alpha = 0.4$ in the following.

From equations (\ref{eq: internal energy of cocoon}) and 
(\ref{eq: total mass of cocoon}), the specific internal energy of
the cocoon, $\epsilon_c$, can be obtained as follows:
\begin{eqnarray}
\frac{\epsilon_c}{c^2} = \frac{1}{\Gamma - 1}\frac{P_c}{\rho_c c^2} = 
\alpha (\gamma_j h_j - 1) = 
\alpha \biggl [\gamma_j \biggl (1+\Gamma \frac{\epsilon_j}{c^2} \biggr ) - 1 \biggr ]\;,
\label{eq: specific internal energy of cocoon}
\end{eqnarray}
where $\rho_c = M_c / V_c$ is the rest-mass density of the
cocoon and $\epsilon_j$ is the specific internal energy of the jet.
The pressure of the jet ``temporal-averagely'' coincides with
that of the cocoon as mentioned above. In addition, needless to say,
there is no pressure jump across the interface between the jet and
cocoon because the jet interface is regarded as a contact discontinuity. 
Therefore, the onset condition for the RTI (\ref{eq: onset condition})
at the jet interface is rewritten, with equation (\ref{eq: specific internal
energy of cocoon}), by 
\begin{eqnarray}
\gamma_j^2 \biggl ( \frac{c^2}{\epsilon_j} + \Gamma^2 \biggr ) >
\biggl ( \frac{c^2}{\epsilon_c} + \Gamma^2 \biggr )
= \biggl ( \frac{1}{\alpha (\gamma_j h_j - 1)} + \Gamma^2 \biggr ) \;.
\label{eq: estimation of onset condition}
\end{eqnarray}
Solving equation (\ref{eq: estimation of onset condition}) numerically,
we can find that the onset condition for the RTI is satisfied regardless
of the magnitude of $\epsilon_j$ when $\gamma_j \gtrsim 1.2$ in 
cases $\Gamma = 4/3$ or $5/3$. Furthermore, when we suppose
the relativistically hot jet ($\epsilon_j /c^2  \gg 1$ and $h_j  \gg 1$) or
cold jet ($\epsilon_j /c^2  \ll 1$ and $h_j \simeq 1$), the onset condition
(\ref{eq: estimation of onset condition}) reduces to 
\begin{eqnarray}
\gamma_j > 1 \;.
\end{eqnarray}
This condition is always fulfilled in the jet. Therefore, overall our
analysis predicts that the onset condition for the RTI at the interface
between the jet and cocoon is satisfied when the jet propagates
through the uniform ambient medium with a sufficient relativistic
velocity ($\gamma \gtrsim 1.2$).

We verify the validity of our prediction in the next section. We discuss,
in Section 4.2., the impact of the pressure/density stratification of the
ambient medium, in which the ambient pressure decays with the jet
propagation, on the growth of the RTI. 
%%%%%%%%%%%%%%%%%%%%%%%%%%%%%%%%%%%%%%%%%%%%%%%%%%%%%
%%%%%%%%%%%%%%%----------S3 ---------%%%%%%%%%%%%%%%
%%%%%%%%%%%%%%%%%%%%%%%%%%%%%%%%%%%%%%%%%%%%%%%%%%%%%
\section{Numerical verification of analytic prediction}
\subsection{Numerical setup}
To quantitatively confirm the analytic prediction presented above, 
we have conducted 3D SRHD simulations of the jet propagating 
through the uniform ambient medium in a cylindrical coordinate 
system $(r, \phi, z)$.  Assuming the ideal-gas law with the ratio
of specific heats $\Gamma = 4/3$, the governing equations are
\begin{eqnarray}
\frac{\partial}{\partial t}(\gamma \rho)                      &+& \nabla \cdot (\gamma  \mbox{\boldmath $v$}) = 0 \;, \label{eq: mass conservation} \\ 
\frac{\partial}{\partial t}(\gamma^2 \rho h \mbox{\boldmath $v$})  &+&  \nabla \cdot (\gamma^2 \rho h \mbox{\boldmath $v$} \mbox{\boldmath $v$} + P {\bf I}) = 0 \;, \label{eq: momentum conservation} \\
\frac{\partial}{\partial t}(\gamma^2 \rho h c^2  - P) &+&  \nabla \cdot (\gamma^2 \rho h c^2  \mbox{\boldmath $v$}) = 0 \;, \label{eq: energy conservation}
\end{eqnarray}
where ${\bf I}$ is the unit matrix and the other symbols maintain
their previous meanings. Note that ``$\nabla \cdot$'' denotes
the divergence of variables in the cylindrical coordinates. 

Additionally, we independently solve the advection of a passive
tracer, which distinguishes the jet material ($f=1$) from the ambient
medium ($f=0$), in conservation form:
\begin{eqnarray}
\frac{\partial}{\partial t}(\gamma \rho f) + \nabla \cdot (\gamma \rho f \mbox{\boldmath $v$}) = 0 \;.
\label{eq: tracer}
\end{eqnarray}

A relativistic HLLC scheme \citep{Mignone05} is used to solve 
equations~(\ref{eq: mass conservation})--(\ref{eq: tracer}) in 
conservation form. The primitive variables are calculated from 
the conservative variables following the method of \citet{Mignone07}. 
We employ a MUSCL-type interpolation method to attain second-order 
accuracy in space, while second-order temporal accuracy is achieved 
using Runge-Kutta time integration. (See Matsumoto et al. 2012
and MM13 for details on our SRHD code.)

Initially, the calculation domain is filled with a homogeneous 
stationary ambient medium. A relativistic cylindrical jet with a 
radius of $r_{j,0}$ and aligned with the $z$-axis is 
continuously injected into the domain from the lower boundary 
at $z$ = 0. The calculation domain spans $0 < r/r_{j,0} < 150$, 
$0 < \phi < 2\pi$ and $0 < z/r_{j,0} < 300$. The normalisation 
units in length, velocity, time and energy density are respectively
selected as the jet radius at the injection point $r_{j,0}$, the speed
of light $c$, the light crossing time over the jet radius $r_{j,0}/c$ 
and the rest-mass energy density in the ambient medium
$\rho_{a}c^2$.

%%%%%%%%%%%%%%%%%%%%--------------------Table 1---------------------------%%%%%%%%%%%%%%%%%%%%
\begin{table*}
\caption{List of simulation runs. The parameters are chosen
so as the four models to be the representatives in parameter
space of $h_{j} - \eta_{j,a}$.}
\begin{tabular}{lcccccclccc}
\hline
Model & dominant energy & $\gamma_{j}$ & $\rho_{j}c^2$ & $P_{j}$ & $h_{j}$ & $\epsilon_{j}/c^2$ & $\eta_{j,a}$ & $L_{j}$ & $\rho_{a}c^2$ & $P_{a}$\\
\hline
H- & internal & $5$ & $10^{-4}$ & $10^{-3}$  & $41$ & $0.03$ & $10^{-1}$ & $10^{-1}\pi$ & $1$ & $10^{-4}$\\
H+ & internal & $5$ & $10^{-2}$ & $10^{-1}$  & $41$ & $30$ & $10^{1}$ & $10^{1}\pi$ & $1$ & $10^{-2}$\\
C- & kinetic & $31$ & $10^{-4}$ & $10^{-6}$  & $1.04$ & $0.03$ & $10^{-1}$ & $10^{-1}\pi$ & $1$ & $10^{-7}$\\
C+ & kinetic & $31$ & $10^{-2}$ & $10^{-4}$  & $1.04$ & $30$ & $10^{1}$ & $10^{1}\pi$ & $1$ & $10^{-5}$
\label{table1}
\end{tabular}
\end{table*}
%%%%%%%%%%%%%%%%%%%%%%%%%%%%%%%%%%%%%%%%%%%%%%%%%%%%%%%%%%%%%

The grid spacing in the $r$-direction is grouped into two regions;
an inner domain ($0<r/r_{j,0}<15$) and outer domain ($15 <
r/r_{j,0}<150$). We use a uniformly spaced grid consisting of 150 
zones in the inner domain, whereas a geometrically stretched 
grid is adopted for the 40 zones of the outer domain. The injected
jet is resolved by 10 numeric cells at the lower boundary and then, 
the main body of the jet is located within the inner domain during
jet propagation for all our simulations. The polar angle of the calculation 
domain is uniformly divided into $\Delta \phi = \pi /80$. We also use
a uniformly spaced grid consisting of 3000 zones in the $z-$direction. 
The grid sizes for the $r$- and $z$- directions in the inner domain
are same; $\Delta r /r_{j,0} = \Delta z/r_{j,0} = 0.1$. While a reflective
boundary condition is imposed on the lower boundary outside the
injection region of the jet ($r/r_{j,0} > 1$), an outflow (zero gradient)
boundary condition is adopted for the outer boundary of the calculation
domain. 

The coordinate singularity is addressed by placing no grid point
on the cylindrical axis and filling appropriate ``ghost grids'' in
the region $r < 0$ \citep{Mohseni00, Ghosh10}. The physical
variables are defined at the grid cell centre. The computational
cell centre ($r_{i}, \phi_{j}, z_{k}$) in the inner domain is defined
as follows;
\begin{eqnarray}
r_{i} &=& \biggl ( i -\frac{1}{2} \biggr) \Delta r, \; (i=1, ... \;, N_{r}) \;, \\
\phi_{j} &=& \biggl (j -\frac{1}{2} \biggr ) \Delta \phi, \; (j=1, ... \;, N_{\phi}) \;, \\
z_{k} &=&  \biggl (k - \frac{1}{2} \biggr ) \Delta z, \; (k=1, ... \;, N_{z}) \;,
\end{eqnarray}
where $(N_{r}, \; N_{\phi}, \; N_{z}) = (150, 160, 3000)$. The physical
variables of the exterior ghost grid at $(r_{0}, \;\phi_{j}, \;z_{k})$ are
assigned to those of the interior grid at $(r_{1}, \; \phi_{j+N_{\phi}/2},\;
z_{k})$ considering the sign as follows;
\begin{eqnarray}
\mbox{\boldmath $q$}_{r_{0}, \; \phi_{j}, \; z_{k}}               &=&   \mbox{\boldmath $q$}_{r_{1}, \; \phi_{j+N_{\phi}/2}, \; z_{k}} \;, \\
\mbox{\boldmath $q$}^{\prime}_{r_{0}, \; \phi_{j}, \; z_{k}} &=& - \mbox{\boldmath $q$}^{\prime}_{r_{1}, \;\phi_{j+N_{\phi}/2}, \; z_{k}} \;,
\end{eqnarray}
where $\mbox{\boldmath $q$} = (\rho, \; P, \; v_{z}, \; f)$ and
$\mbox{\boldmath $q$}^{\prime} = (v_{r}, \; v_{\phi})$.
When we consider the second-order accuracy in space,
\begin{eqnarray}
\mbox{\boldmath $q$}_{r_{-1}, \; \phi_{j}, \; z_{k}}               &=&   \mbox{\boldmath $q$}_{r_{2}, \; \phi_{j+N_{\phi}/2}, \; z_{k}} \;, \\
\mbox{\boldmath $q$}^{\prime}_{r_{-1}, \; \phi_{j}, \; z_{k}} &=& - \mbox{\boldmath $q$}^{\prime}_{r_{2}, \;\phi_{j+N_{\phi}/2}, \; z_{k}} \;.
\end{eqnarray}
When using the physical variables at the exterior ghost grids,
we evaluate the numerical fluxes at the inner-most computational
grids for the HLLC scheme.
%%%%%%%%%%%%%%%%%%%%--------------------Figure 2---------------------------%%%%%%%%%%%%%%%%%%%%
\begin{figure*}
\begin{center}
\scalebox{0.45}{{\includegraphics{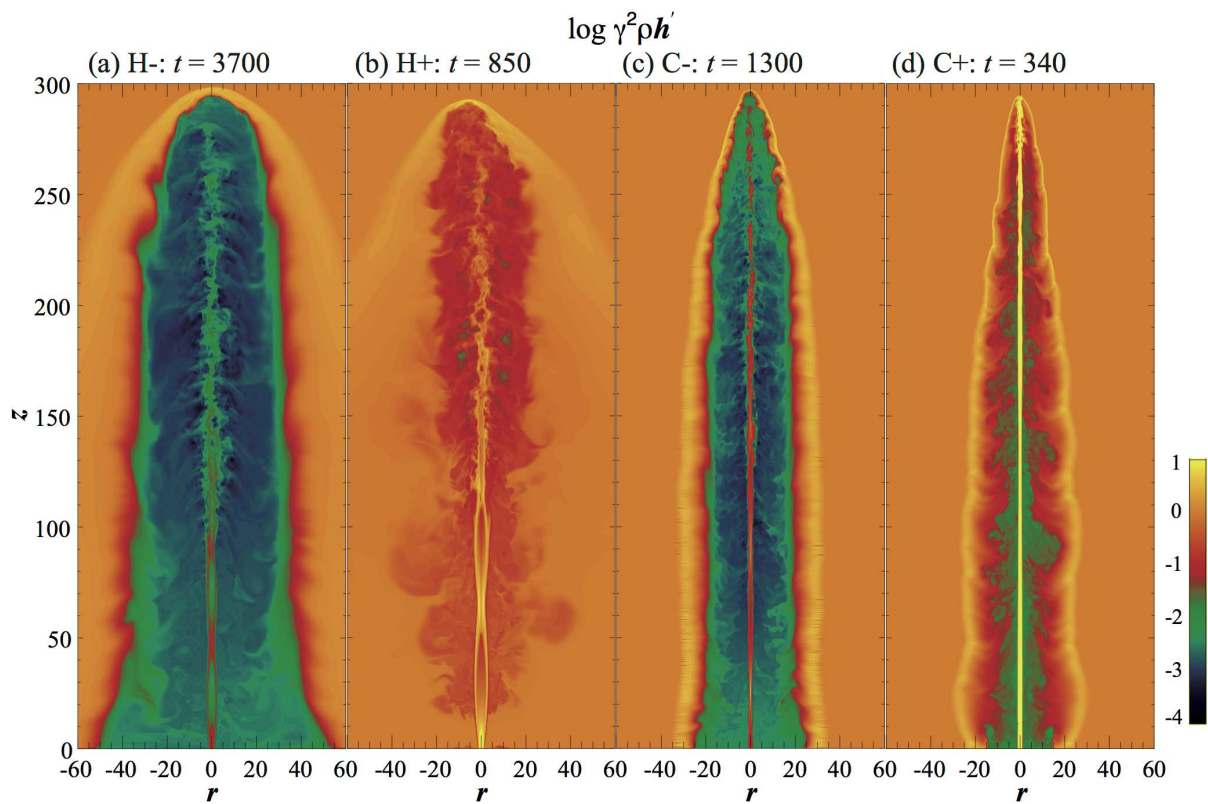}}} 
\caption{Spatial distribution of $\gamma^2 \rho h^{\prime}$ on
the cutting plane along the $z$-axis when the jet head reaches
the upper boundary. Panels~(a), (b), (c) and (d) correspond to
the models H-, H+, C- and C+, respectively.}
\label{fig2}
\end{center}
\end{figure*}
%%%%%%%%%%%%%%%%%%%%%%%%%%%%%%%%%%%%%%%%%%%%%%%%%%%%%%%%%%%%%

\subsection{Jet models}
In this paper we focus on two parameters: relativistic hotness
of the jet, $h_{j}$, and the effective inertia ratio of the jet to the
ambient medium, $\eta_{j,a}$. Here, $\eta_{j,a}$ controls the
morphology and dynamics of propagating jet and is given by
a functional form as will be shown in what follows. Neglecting 
the multi-dimensional effect, the propagation velocity of the jet 
head through the ambient medium, $v_{h}$, can be evaluated 
by balancing the momentum flux of the jet and ambient 
medium in the frame of the jet head as follows:
\begin{eqnarray}
v_{h} = \frac{\sqrt{\eta_{j,a}}}{1+\sqrt{\eta_{j,a}}}v_{j} \label{eq: propagation velocity}
\end{eqnarray}
\citep{Marti97, Mizuta04}. The relativistic internal energy and/or 
Lorentz factor of the fluid enhance the inertia. The effective 
inertia ratio, $\eta_{j,a}$, is thus obtained as
\begin{eqnarray}
\eta_{j,a} = \frac{\gamma_{j}^2 \rho_{j}h_{j}}{\rho_{a}} \;.
\end{eqnarray}
Equation~(\ref{eq: propagation velocity}) 
indicates that the propagation velocity of the jet head, $v_{h}$, 
is roughly equal to the fluid velocity of the relativistic jet, $v_{j}$,
in the regime where $\eta_{j,a} \gg 1$, while $v_{h}$ is much 
slower than $v_{j}$ and is not relativistic in the regime where 
$\eta_{j,a} \ll 1$.

Bearing in mind the confirmation of our prediction in various
jet launching conditions, four-types of models, which are the 
representatives covering the parameter space of $\eta_{j,a}$
and $h_{j}$, are simulated. The parameters adapted in
each model is summarized in Table~{\ref{table1}}.
The relativistic hotness of the jet is distinguished by the labels
``H'' (hot with $h_{j}=41$) or ``C'' (cold with $h_{j}=1.04$).
The relative inertia between jet and ambient medium is
distinguished by the labels containing ``+'' (high relative
inertia with $\eta_{j,a}=10$) or ``-'' (low relative inertia with
$\eta_{j,a}=0.1$).

The relativistic hotness is related to the dominant type of energy
in the jet. The specific internal energy is a good indicator for the
dominant type of energy \citep[e.g.,][]{Marti16}. The internal energy
is dominant in the regime where $\epsilon/c^2 > 1$ while the kinetic
energy is dominant in the regime where $\epsilon/c^2 < 1$. The
relation between the relativistic hotness, $h$, and the specific internal
energy, $\epsilon$, is given by $h = 1 + \Gamma \epsilon/c^2$.
The specific internal energy of the jet, $\epsilon_{j}/c^2$, in
the hot and cold cases are $30$ and $0.03$, respectively.
This indicates that hot and cold models in our simulations
are classified as internal energy dominated and kinematically
dominated, respectively.

The energy flux of the jet is defined by
\begin{eqnarray}
L_{j} = \pi r_{j,0}^2 \gamma_{j}^2 \rho_{j} h_{j} c^2 v_{j} \sim \pi \eta_{j,a} r_{j,0}^2 \rho_{a} c^3 \;.
\end{eqnarray}
Therefore, the jet power of the large inertia model (H+ and C+)
is $10 \pi$ while that of the small inertia model (H- and C-) is $0.1 \pi$.
When we assume that the size of the jet radius at the nozzle is 
about $10$ pc, the energy flux of the jet is estimated, for the case
of the jet with small $\eta_{j,a}$, as follows;
\begin{eqnarray}
L_{j}  & \sim &10^{44} \; {\rm erg/s} \; \nonumber \\
&\times& \biggl  ( \frac{r_{j,0}}{10 \; {\rm pc}} \biggr )^2 
\biggl ( \frac{\gamma_{j}}{5} \biggr ) ^2 
\biggl ( \frac{\rho_{j}/\rho_{a}}{10^{-4}} \biggr ) 
\biggl ( \frac{\rho_{a}}{1.4 \times 10^{-26} \; {\rm g/cm^3}} \biggr ) 
\biggl ( \frac{h_{j}}{41} \biggr ) \; ,
%\biggl ( \frac{v_{j}}{0.98c} \biggr ) \;.
\end{eqnarray}
where we chose $\rho_{a}$ as the typical ambient density in
the AGN system of $10^{-26}$ g/cm$^3$.
If we suppose the jet with large $\eta_{j,a}$, $L_{j}$ is estimated
as $10^{46}$ erg/s. In the case of a GRB jet propagating through
a progenitor star with $r_{j,0} \sim 10^{7}$ cm and $\rho_a \sim 10^{5}$
g/cm$^3$, $L_{j}$ of small (large) inertia models are estimated as
$10^{50}$ erg/s ($10^{52}$ erg/s).

In our simulation models, the pressure ratio between the jet
and ambient at the jet nozzle is set to be a relatively large value,
i.e., $P_{j}/P_{a} = 10$. In this setting, the hot jet (models H- and
H+) starts its evolution from the expansion stage, whereas the
cold jet (models C- and C+) starts its evolution from the contraction
stage because of the generation of the overpressured cocoon.
If we change the initial pressure ratio, it does not have a significant
impact on our results because the inertia ratio between the jet 
and ambient medium and the hotness of the jet which are the chosen
key parameters in our simulations are mainly responsible for the
morphology of the cocoon \citep{Marti97}. Actually, recent 3D simulations 
of the AGN jet propagation reported that morphologies between the case
with $P_{j}/P_{a} = 10$ and $1$ were quite similar \citep{Li18}.

A small-amplitude ($1$\%) random pressure perturbation is
introduced to the injected relativistic flow to break the symmetry
of the jet as follows: White noise, $\delta(t, \; r, \; \phi, \; z)$, is
generated at every time step for the pressure at the jet nozzle
($r < r_{j,0}$ and $z < 0$). The maximum and minimum value of
$\delta(t, \; r, \; \phi, \; z)$ are $0.01$ and $-0.01$, respectively.
The pressure of the jet at the jet nozzle is given by $P_{j}
(1+ \delta(t, \; r, \; \phi, \; z))$ as the boundary condition.
%%%%%%%%%%%%%%%%%%%%%%%%%%%%%%%%%

\subsection{Confirmation of our analytic prediction}
As can be observed, Fig.~\ref{fig2} presents a spatial distribution
of $\gamma^2 \rho h^{\prime}$, which is related to the onset
condition for RTI, on the cutting plane along the $z$-axis at the
final state of the simulation run (that is, when the jet head reaches 
the upper boundary at $z = 300$). Panels~(a), (b), (c) and (d) in
the figure correspond to models H-, H+, C- and C+, respectively. 
Note that the time of the final state ($=t_{\rm end}$) differs 
owing to the propagation velocity of the jet among the four models, 
that is, $t_{\rm end} = 3700$ (H-), $850$ (H+), $1300$ (C-) and 
$340$ (C+).

The propagation velocity of the cold jet is typically faster than 
that of the hot jet when compared between models with the same 
$\eta_{j,a}$. This result can be physically explained by the nonlinear 
dynamics of jet propagation. In the cold model, the pressure 
of the injected jet is generally smaller than that of the cocoon 
envelope because a large amount of the cold jet's kinetic energy 
is converted into the thermal energy of the cocoon at 
the jet head. The initially-cold jet is thus compressed by the 
external cocoon's pressure as time advances and then gains
the larger effective inertia than its hot counterpart. As a result, 
the propagation velocity increases more in the cold jet than 
in the hot jet at the same initial $\eta_{j,a}$. The thickness of 
the cocoon envelope relative to the jet radius also varies 
among the four models. The slower the propagation velocity of
the jet head, the thicker the cocoon envelope. These properties
of the cocoon are common and well known in numerical simulations
of jet propagation \citep[e.g,][]{Marti97, Komissarov97}.

As analytically predicted in Section~2, the $\gamma^2 \rho h^{\prime}$
of the jet becomes larger than that in the cocoon envelope 
at the jet interface for all the models at the final state, regardless 
of the jet launching condition. Since the condition (\ref{eq: onset 
condition}) is satisfied at the interface between the jet and 
cocoon, oscillation-induced RTI can grow there.

\begin{figure}
\begin{center}
\scalebox{0.35}{{\includegraphics{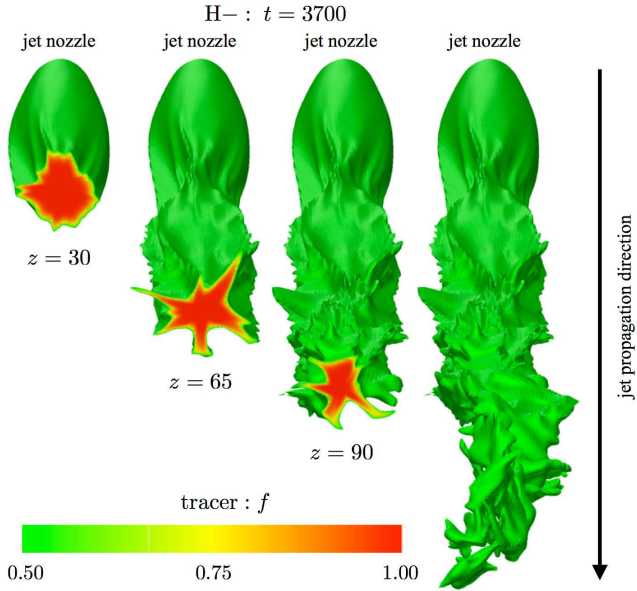}}} 
\caption{3D rendering of the tracer (but only $f>0.5$) at $t=3700$ 
for model H-. The jet propagates from the top to bottom. The red
tone denotes higher tracer value. Note that this plot is not linear
in distance from the top to bottom because the viewing angle is
chosen to exhibit the cross-section of the jet clearly.
}
\label{fig3}
\end{center}
\end{figure}

Fig.~\ref{fig3} presents the 3D rendering of the tracer in the range 
$f > 0.5$ for model H- when $t = 3700$. The rightmost depiction
corresponds to the entire structure. The cross-sections at $z = 30$, 
$65$ and $90$ are also demonstrated for reference. The red tone 
denotes higher tracer value. Because the jet has a larger pressure
compared to the cocoon, the jet expands radially around the
jet nozzle, resulting in the excitation of reconfinement shock inside
the jet. The distortion of the jet interface is clearly observed when
the jet contracts radially (see the jet cross-section at $z=30$ in
Fig.~\ref{fig3}). The growth of the oscillation-induced RTI is
expected to contribute to the formation of the corrugated jet interface.
The possible development of the other instabilities except RMI
is discussed in Section 4.1.

The excited reconfinement shock inside jet converges to the jet
axis and then transforms into radially outward propagating shock. 
When the outgoing shock encounters the contact discontinuity,
RMI grows at the jet interface. Unlike oscillation-induced RTI ,
RMI grows impulsively when the reconfinement shock collides
with the contact discontinuity (see Fig.~\ref{fig1} or MM13 for details
on the excitation mechanism). The growth of the RMI secondary
contributes to the amplification of the amplitude of the corrugated
jet interface and the excitation of the elongated finger-like structures
on the jet cross-section \citep{Matsumoto13}.

It is important to note that, in model H-, the first growth of RMI occurs
around $z = 50$. Oscillation-induced RTI would be responsible
for the distortion of the cross-section at $z = 30$. On the other hand, 
the elongated finger-like structures that appear in the cross-section
at $z = 65$ and $90$ result from the combination of oscillation-induced
RTI and RMI.

\begin{figure}
\begin{center}
\scalebox{0.35}{{\includegraphics{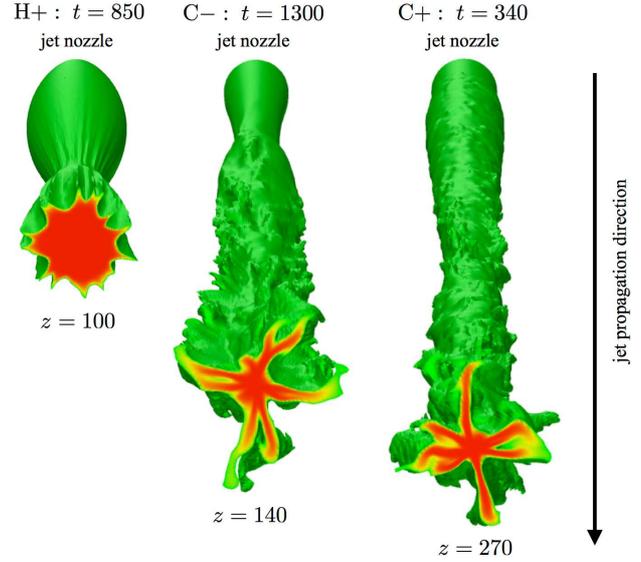}}} 
\caption{3D rendering of the tracer (but only $f>0.5$) for models H+,
C- and C+ when the jet head reaches the upper boundary ($z=300$).
Color means the same as Figure~\ref{fig3}. The red tone denotes
higher tracer value. Jets propagate from the top to bottom. The viewing
angle is chosen to be suited for each model.
}
\label{fig4}
\end{center}
\end{figure}

Fig.~\ref{fig4} illustrates the 3D rendering of the tracer in the range 
of $f > 0.5$ for models H+, C- and C+ just after the jet head 
arrives at the upper boundary. The cross-section for each model 
is also pictured to clearly demonstrate the deformed jet interface. 
The viewing angle is chosen to suit each model respectively. 
In all three models, the elongated finger-like structures are also
excited on the jet cross-section with the radially oscillating motion.
Although the relativistic jet demonstrates a rich variety of propagation 
dynamics depending on the launching conditions, the oscillation-induced 
RTI and associated RMI could grow mutually at the jet interface, 
inducing a significant number of finger-like structures.
Overall, the simulation results confirm our analytic prediction and
suggest that the deformation of the relativistic jet interface
is unavoidable in a purely hydrodynamic regime unless some 
constricting effects exist on the radial oscillating motion of the jet.
The possible effects of the suppression of the growth of the
oscillation-induced RTI are discussed in the following section.
%%%%%%%%%%%%%%%%%%%%%%%%%%%%%%%%%%%

%%%%%%%%%%%%%%%%%%%%--------------------Figure 3---------------------------%%%%%%%%%%%%%%%%%%%
%%%%%%%%%%%%%%%%%%%%%%%%%%%%%%%%%%%%%%%%%%%%%%%%%%%%%%%%%%%%%

%%%%%%%%%%%%%%%%%%%%--------------------Figure 4---------------------------%%%%%%%%%%%%%%%%%%%
%%%%%%%%%%%%%%%%%%%%%%%%%%%%%%%%%%%%%%%%%%%%%%%%%%%%%%%%%%%%%

%%%%%%%%%%%%%%%%%%%%%%%%%%%%%%%%%%%%%%%%%%%%%%%%%%%%%
%%%%%%%%%%%%%%%----------S4 ---------%%%%%%%%%%%%%%%
%%%%%%%%%%%%%%%%%%%%%%%%%%%%%%%%%%%%%%%%%%%%%%%%%%%%%
\section{Discussion}
\subsection{Possible development of instabilities in jets}
Because the effects of the magnetic field and rotation (i.e., the bulk
azimuthal velocity) of jets are ignored in this work for simplicity, the
growth of the CDI and rotation-induced instability \citep{Meliani07,
Meliani09, Millas17} is not expected in this study. 

The development of the KHI in relativistic jets has been investigated
diligently \citep[e.g.][]{Turland76, Blandford76, Ferrari78, Hardee79,
Hardee98, Hardee01, Perucho04, Perucho05, Perucho07a, Mizuno07,
Rossi08, Perucho10}. From the stability analysis of jets to the KHI,
it is widely known that, in the case the radial oscillating motion of the jet
is absent, the reflection body mode becomes dominant when the flow
velocity of a jet is supersonic \citep[e.g.,][]{Payne95, Perucho04,
Perucho10}. This mode invokes the small-scale disturbance of the 
jet interface while the surface mode contributes to the global-scale
deformation of the jet. Therefore, even in our models, the body mode
of the KHI may play a role in generating the small-scale disturbances
at the jet interface. 

Not only the RTI and RMI, CFI should develop as well when the jet has
radial oscillating motion \citep{Gourgouliatos18a, Gourgouliatos18b}.
The jet moves along curved streamlines with relativistic velocity at the
radially oscillating interface of the jet (blue lines in Fig.\ref{fig1}), where
the centrifugal force essentially balances the pressure gradient force of
the jet. However, the cocoon does not move fast at the jet interface
compared to the jet. Therefore, the angular velocity along the curved
streamlines jumps across the jet interface and drastically decreases
from the jet towards the cocoon. The onset condition for the growth of
CFI is thus expected to be satisfied in this situation \cite[see,][for details]
{Gourgouliatos18a}.

Since the inertia force, which induces the oscillation motion of
the jet, is responsible for both the RTI and CFI, the typical growth
time of these instabilities is expected to be compatible with the
oscillation time-scale of the jet, $\tau_{\rm osci}$. Qualitatively,
the $\tau_{\rm osci}$ can be estimated as the free-fall time of the
jet radius, that is $\tau_{\rm osci} \sim \sqrt{r_j/g}$ where $r_j$ is
the jet radius and $g$ is an effective acceleration. In the case of
the RTI, above speculation is consistent with the linear theory
\citep{Matsumoto17}. The typical linear growth time of the RTI is
given by $\tau_{\rm RTI} \sim 1/\sqrt{gk}$ where $k$ is the wavenumber.
When taking $k \sim 1/r_j$, we can find that $\tau_{\rm RTI}$
coincides with $\tau_{\rm osci}$. 

Since the origin of oscillation-induced instabilities is the same and
no difference in the growth rate of the instabilities is expected,
it is difficult to distinguish the RTI and CFI at a glance. However,
it is obvious that only the CFI grows at the jet interface in the case
where the jet is surrounded by a dense medium compared to the jet
\citep{Gourgouliatos18a} because the jet is RTI stable in such a situation. 

\begin{figure}
\begin{center}
\scalebox{0.45}{{\includegraphics{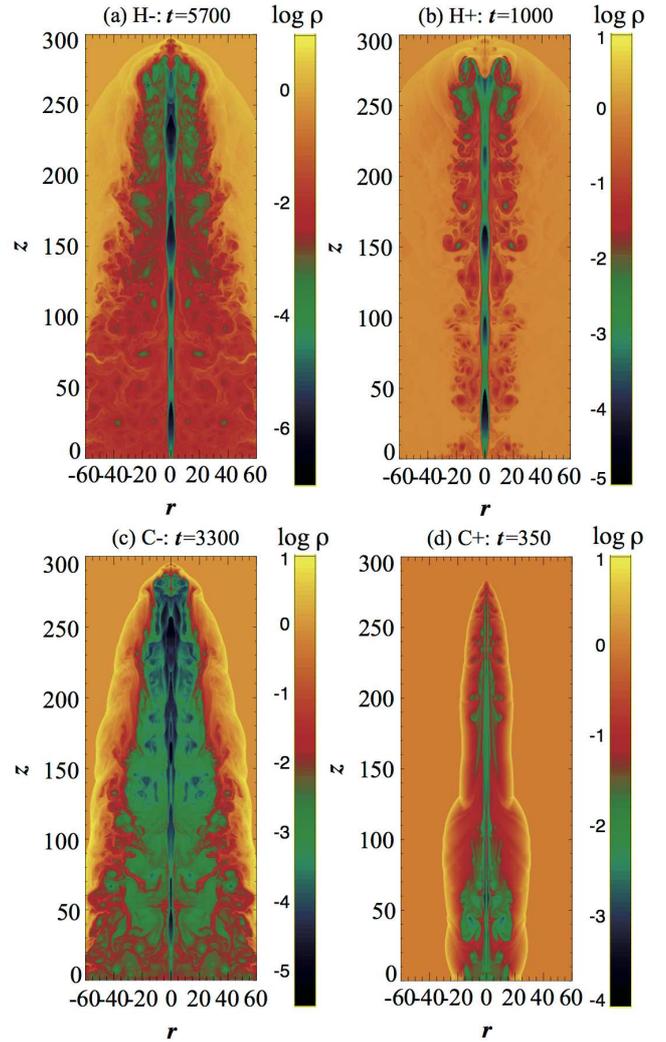}}} 
\caption{Density distribution of 2D axisymmetric simulations
for four models (H-, H+, C- and C+).}
\label{fig5}
\end{center}
\end{figure}

As described in the introduction, the non-axisymmetric modes of
the RTI and RMI cannot be excited in conventional axisymmetric
simulations of jet propagation. In order to confirm it, 2D
counterparts for all the four models are simulated in the cylindrical
coordinate system with the same resolutions as 3D models, that is,
$\Delta r /r_{j,0} = \Delta z/r_{j,0} = 0.1$. Fig.~\ref{fig5} shows the 
density distribution of 2D counterparts. As expected, we can find
that the jet can propagate stably under the constraint of the 
axisymmetry. The results of these 2D counterparts indicate that the
non-axisymmetric modes of some instabilities are responsible for
the destabilization of the jet in our 3D simulations although it is still 
difficult to identify the main player. It would be a challenge for our
future work to identify the dominant instability mode.

\subsection{A simple relationship between hotness and velocity of the jet and its stability}
A simple scaling relation between the hotness and velocity of the jet
and its stability to the RTI can be obtained from a qualitative argument. 
The inertia force, which induces the radially oscillating motion of the jet,
balances with the pressure gradient force of the jet: 
\begin{eqnarray}
\gamma_j^2 \rho_j h_j g = - \nabla P_j \;.
\end{eqnarray}
This relation provides a rough estimation of the acceleration $g$ as 
\begin{eqnarray}
g \sim \frac{c_{s,j}^2}{\gamma_j^2 r_j} \;,
\end{eqnarray}
where $c_{s,j}$ is the sound speed of the jet. Therefore, the growth
rate of the oscillation-induced RTI, $\sigma$, has the following relation; 
\begin{eqnarray}
\sigma \sim \frac{1}{\tau_{osci}} \propto \sqrt{\frac{g}{r_j}} 
\sim \frac{c_{s,j}}{\gamma_j r_j} \sim \frac{v_j}{\gamma_{s,j} r_j} \frac{1}{M_{rela, j}} \; , \label{eq: growth rate osci}
\end{eqnarray}
where $M_{rela, j} = \gamma_j v_j / \gamma_{s,j} c_{s,j}$ is the relativistic
Mach number of the jet \citep{Konigl80, Komissarov98} and $\gamma_{s, j}$
is the Lorentz factor of the jet’s sound speed. This indicates that the hotter jet
is more unstable to the RTI than the colder jet because the growth rate of it
is proportional to the sound speed of the jet. In addition, the faster (higher
relativistic Mach number and/or Lorentz factor) jet becomes more stable to
the RTI. These tendencies of the jet stability to the RTI in the linear regime
are similar to those to the KHI \citep{Perucho10} and are observable 
even in our simulations when comparing the jet stability between the
models with the same inertia. Hot jets lose their coherent shapes violently
compared to their cold and faster counterparts as shown in Fig.~\ref{fig2}.
%%%%%%%%%%%%%%%%%%%%--------------------Figure 5---------------------------%%%%%%%%%%%%%%%%%%%
%%%%%%%%%%%%%%%%%%%%%%%%%%%%%%%%%%%%%%%%%%%%%%%%%%%%%%%%%%%%%

\subsection{Jet propagation through decaying pressure medium}
In the actual astrophysical system, the jet propagates through
an inhomogeneous medium where the density and pressure
drop with the propagation \citep[e.g.,][]{Perucho07b, Perucho11}.
Then, the thermodynamical properties of the cocoon are expected
to change with time. Since the physical properties, such as the
typical growing scale, of oscillation-induced instabilities are associated
with the reconfinement of the jet by the cocoon, a change of the
propagation environment should impact on the stability of the jet. 

The reconfinement region and thus the jet radius become large with
time when the jet propagates through the ambient gas with a decaying
density and pressure profiles \citep[e.g.,][]{Matsumoto12}. In such
a situation, the growth rate of the oscillation-induced instabilities becomes
small compared to the case where the jet propagates through the uniform
ambient medium. The larger jet radius provides the longer oscillation time.
This would be the reason for it. We note that the rough estimation of
the growth rate of the oscillation-induced RTI (\ref{eq: growth rate osci})
also predicts such a behavior of the jet though it is derived from the local
balancing of forces. 

We finally stress that, when the ambient density and pressure decrease
drastically and the jet enters into the free expansion phase, the jet
becomes stable to the oscillation-induced instabilities because in such
a situation, the reconfinement process no longer works, resulting in
a loss of causal connectivity across the jet \citep{Porth15}. 

\subsection{Effect of magnetic field} 
The magnetic field plays a variety of roles in the propagation of relativistic
jets \citep[e.g.,][]{Komissarov99, Leismann05, Bromberg14, Mizuno15, Marti16}.
The toroidal component of it pinches the jet and weakens the inward motion
of the rarefaction waves, resulting in the formation of weaker reconfinement
shocks and rarefaction waves than those in a purely HD jet. On the other
hand, since the axial component of it gives rise to a magnetic pressure
additional to the gas pressure, it leads to stronger reconfinement shocks
and rarefactions. 

The toroidal magnetic field is expected to be dominant in the jet
at the larger distance from the jet formation region \citep{Baum97,
Laing14}. When considering a strong-ordered magnetic field, the
magnetic tension force due to the toroidal field would more or less
contribute to the suppression of non-axisymmetric modes of the
oscillation-induced RTI which have the wavevector parallel to the
direction of the magnetic field.

In the non-relativistic regime, when the wavevector, $ \mbox{\boldmath $k$}$,
is parallel to the magnetic field, $ \mbox{\boldmath $B$}$, the growth rate of
the magnetic RTI is given by 
\begin{eqnarray}
\omega^2 = -gk \biggl [ A - \frac{B^2 k}{2\pi (\rho_{+} + \rho_{-})g } \biggr ] \;, \label{eq: growth rate MRTI}
\end{eqnarray}
where $\rho_{+}$ and $\rho_{-}$ are the upper and lower
density against an acceleration, $g$, respectively, and
$A = (\rho_{+} - \rho_{-})/(\rho_{+} + \rho_{-})$ is the Atwood number
\citep{Chandrasekhar61, Hillier06}. At the limit $B = 0$, this reduces
to the growth rate of the RTI for purely hydrodynamic case;
\begin{eqnarray}
\omega^2 = -gkA \;.
\end{eqnarray}
The critical wavelength $\lambda_{c}$ is obtained from above dispersion
relation for the magnetic RTI as 
\begin{eqnarray}
\lambda_{c} = \frac{B^2}{(\rho_{+} - \rho_{-})g} \;. \label{eq: critical wavelength}
\end{eqnarray}
The strength of the magnetic field characterizes $\lambda_c$ and the RTI
mode with $\lambda < \lambda_c$ is then suppressed due to the magnetic
tension force.

Next, the magnetic RTI in the relativistic regime is in our interest. However,
it is not sufficiently investigated even at the linear stage. Here we speculate
the linear property of the relativistic magnetic RTI in analogy with its non-relativistic
counterpart. 

The growth rate of the relativistic RTI for purely hydrodynamical case is given by
\begin{eqnarray}
\omega^2 = -gk \mathcal{A} \;, \label{eq: relativistic dispersion relation}
\end{eqnarray}
where $\mathcal{A}$ is the relativistic Atwood number \citep{Matsumoto17}.
Based on this, the growth rate and the critical wavelength of the ``relativistic''
magnetic RTI in the jet--cocoon system are surmised as follows: The density
in equations~(\ref{eq: growth rate MRTI}) and~(\ref{eq: critical wavelength})
would be replaced, physically and intuitively, by the effective inertia in the
relativistic regime. Based on the dispersion relation for the relativistic
hydrodynamic RTI (\ref{eq: relativistic dispersion relation}), the Atwood number
in equation (\ref{eq: growth rate MRTI}) should be replaced by the relativistic
Atwood number as well. Then, the growth rate and critical wavelength of the
relativistic magnetic RTI at the jet interface are expected to be
\begin{eqnarray}
\omega^2 \sim -gk \biggl [ \mathcal{A} - \frac{B^2 k}{2\pi (\gamma_{j}^2 \rho_{j} h_{j} + \gamma_{c}^2 \rho_{c} h_{c})g } \biggr ] \;,
\end{eqnarray}
and
\begin{eqnarray}
\lambda_{c} \sim \frac{B^2}{(\gamma_{j}^2 \rho_{j} h_{j} - \gamma_{c}^2 \rho_{c} h_{c})g} \;,
\end{eqnarray}
respectively.

When the inertia of the jet is much larger than the cocoon, the RTI grows faster.
Even in the limit where $\gamma_{j}^2 \rho_{j} h_{j} \gg \gamma_{c}^2 \rho_{c} h_{c}$,
the critical wavelength is not vanished and continues to exist. In such a situation,
when considering the balance between the pressure gradient force and inertia
force at the jet interface, the typical wavelength of the magnetic RTI is given,
with the similar procedure as Section 4.2, by
\begin{eqnarray}
\lambda_{c} \sim \frac{8\pi r_j}{\beta} \;,
\end{eqnarray}
where $\beta = P_j /(B^2/8\pi)$ is the plasma beta. Since the diameter of the
jet cross-section is given by $2\pi r_j$, all the modes of the RTI potentially
excited in the jet would be suppressed if $\beta$ is the order of or smaller
than unity.

Not only the RTI, the magnetic field also gives various impacts on the instabilities
and their resultant in the jet. For example, \citet{Komissarov19} show recently
that the magnetic tension force due to the toroidal field contributes to the suppression
of the oscillation-induced CFI if the strength of it is greater than a certain level.
When the strength of the magnetic field is weaker, the oscillation-induced
instabilities are expected to amplify the magnetic field via small-scale turbulent
dynamo.

In contrast, there exists a case that the magnetic field itself becomes unstable.
When the jet is strongly dominated by the azimuthal magnetic component, it is well
known that the jet becomes unstable to CDI \citep[e.g.,][]{Eichler93, Mizuno09, Porth15}.
Overall the impact of the magnetic field on the jet propagation dynamics is still
controversial because it is deeply related to the origin of it, in other word, the central
engine of the jet. It is not fully explored and within the scope of our future work. 
%%%%%%%%%%%%%%%%%%%%%%%%%%%%%%%%%%%
\section{Summary}
The nonlinear stability of relativistic jets propagating through the uniform 
ambient medium was studied analytically and numerically by considering
the simultaneous formation of cocoons. Our analytic findings indicate that
the onset condition for oscillation-induced RTI is satisfied at the interface
between a fully relativistic jet and its cocoon, regardless of launching
conditions when the jet oscillates radially. To verify this prior expectation,
we performed 3D SRHD simulations of a relativistic jet propagating through
the homogeneous ambient medium. Based on the parameters studied with
two varying fundamental jet launching conditions (hotness of the jet and
effective inertia ratio between the jet and ambient medium), we confirmed
that oscillation-induced RTI also grows at the relativistic jet interface in
addition to RMI, oscillation-induced CFI and KHI. Hence, we can conclude
that the synergetic growth of these instabilities at the jet interface is an
inherent property of relativistic hydrodynamic jets propagating through
the uniform medium. 

To correctly capture HD and MHD instabilities and obtain the whole
picture of a relativistic jet in its actual astrophysical environment,
high-resolution 3D MHD modelling is required. We have attempted
the first step towards fully understanding astrophysical relativistic
jets in this study. 
%%%%%%%%%%%%%%%%%%%%%%%%%%%%%%%%%%%%%%%%%%%%%%%%%%%%%%%%%%%%%%%%%%%%%%%%%%%%%%%%%%%%%%
\section*{Acknowledgements}
We thank H. R. Takahashi, A. Mizuta, S. Nagataki, M. A. Aloy, J. M. Mart{\'{\i}}, 
M. Perucho, S. S., Komissarov and K. N. Gourgouliatos for their useful 
discussions. Numerical computations were conducted on Cray XC30 
at the Center for Computational Astrophysics, National Astronomical 
Observatory of Japan and on Cray XC40 at YITP at Kyoto University. 
This work was supported in part by Research Institute of Stellar 
Explosive Phenomena at Fukuoka University and the Center for the 
Promotion of Integrated Sciences (CPIS) of Sokendai.
This work was supported by JSPS KAKENHI Grants No. 18K03700,
No. 18H04444, No. 18H01212 and No. 19K23443.
%%%%%%%%%%%%%%%%%%%%%%%%%%%%%%%%%%%%%%%%%%%%%%%%%%%%%%%%%%%%%%%%%%%%%%%%%%%%%%%%%%%%%%

\bibliographystyle{mnras}
\bibliography{./papers}

\appendix

\section{Resolution study in growth of oscillation-induced instabilities in model H+}
%%%%%%%%%%%%%%%%%%%%--------------------Figure A1---------------------------%%%%%%%%%%%%%%%%%%%
\begin{figure*}
\begin{center}
\scalebox{0.5}{{\includegraphics{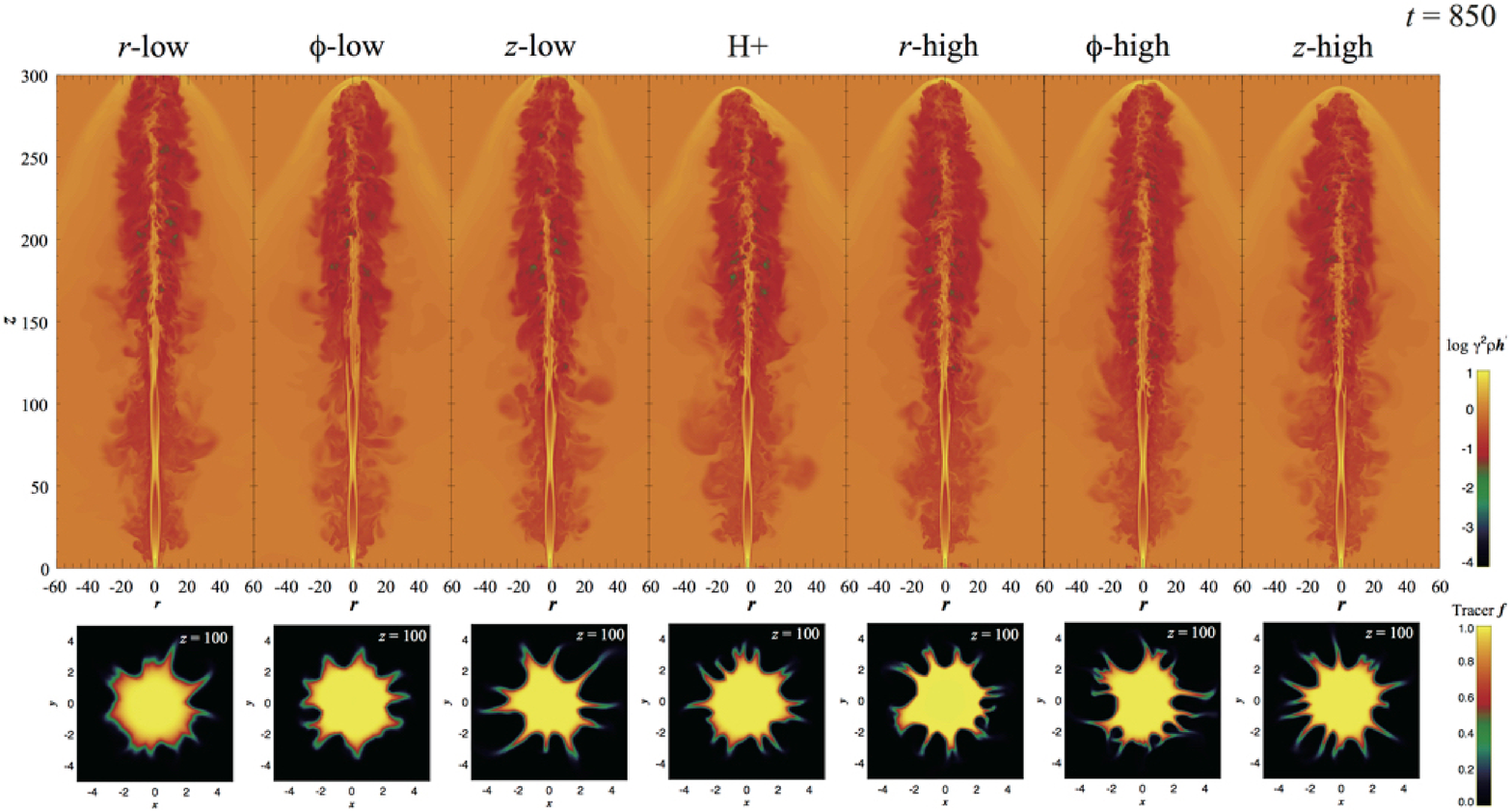}}} 
\caption{Upper panels: Spatial distribution of $\gamma^2 \rho h^{\prime}$
on the cutting plane along the $z$-axis at $t=850$. Lower panels: 
2D cut of the passive tracer in the $x$-$y$ plane at $z=100$ when $t=850$.}
\label{figA1}
\end{center}
\end{figure*}
%%%%%%%%%%%%%%%%%%%%%%%%%%%%%%%%%%%%%%%%%%%%%%%%%%%%%%%%%%%%%
The influence of the numerical resolution on the stability of the jet
is investigated by changing the grid spacing for one particular
simulation model. For this purpose, the model H+ is chosen as
a reference because the evolving finger-like structure, which is
one of the typical features of the oscillation-induced instabilities,
is the most remarkable in this model compared with the others
(i.e, H-, C+ and C-).

The simulation setup of model H+ was described in Section 3.1.
The resolution of the fiducial run is $(N_{r}, \; N_{\phi}, \; N_{z})
= (150, 160, 3000)$. We run $6$ additional models only with
changing the grid spacing and compare them with the fiducial one.
The number of grid points in each direction is summarized for
each model in table~\ref{tableA1}. The direction in which we
change the resolution is distinguished by the labels ``$r$'',
``$\phi$'' and ``$z$'' in the model name. The label containing
``high (low)'' is corresponding to the model with the twice (half)
of the directional resolution adopted in the fiducial one. 

The upper panels of Fig.~\ref{figA1} are the distributions of
$\gamma^2 \rho h^{\prime}$ on the cutting plane along the $z$-axis
at $t = 850$ for the models with different resolutions. In all the models,
$\gamma^2 \rho h^{\prime}$ of the jet is larger than that of cocoon
at the jet interface, indicating that the onset condition for the
oscillation-induced RTI is satisfied. As a result of it, we can find that
the jet is deformed in all the models regardless of the number of
grid points. Shown in the lower panels of Fig.~\ref{figA1} is the
distribution of the passive tracer, $f$, on the $x$-$y$ cutting plane
at $z = 100$ when $t = 850$ for each model. We can find that
the finger-like structure is excited and evolved regardless of
the number of the grid points in all the models, while the typical
azimuthal wavelength of it is different between models and seems
to be corresponding to $\sim 10$ numeric cells for each model.
This implies that, the higher the numerical resolution, the more
the jet is destabilized by shorter wavelength modes with higher
growth rate.

As a conclusion, overall results of our resolution study indicate that
the grid spacing adopted in our simulation study is enough, at least,
to discuss the stability of the jet qualitatively. However, it may be
necessary to use much more numerical grids to study ``quantitatively''
the magnitudes of the mixing, transport, and their impacts on the jet
propagation dynamics at the nonlinear stage. 
%%%%%%%%%%%%%%%%%%%%--------------------Table A1---------------------------%%%%%%%%%%%%%%%%%%%%
\begin{table}
\caption{Number of grid points in each model for the resolution study.}
\begin{tabular}{lccc}
\hline
Model & $N_{r}$ & $N_{\phi}$ & $N_{z}$ \\
\hline
H+ & 150 & 160 & 3000 \\
$r$-low & 75 & 160 & 3000 \\
$\phi$-low & 150 & 80 & 3000 \\
$z$-low & 150 & 160 & 1500 \\
$r$-high & 300 & 160 & 3000 \\
$\phi$-high & 150 & 320 & 3000 \\
$z$-high & 150 & 160 & 6000 \\
\label{tableA1}
\end{tabular}
\end{table}
%%%%%%%%%%%%%%%%%%%%%%%%%%%%%%%%%%%%%%%%%%%%%%%%%%%%%%%%%%%%%

\label{lastpage}

\end{document}